\documentclass[9pt,technote]{IEEEtran}
\usepackage{algorithm}
\usepackage{algorithmic}
\usepackage{amsfonts}
\usepackage{amsmath}
\usepackage{amssymb}
\usepackage{amsthm}
\usepackage{array}
\usepackage{bigints}
\usepackage{booktabs}
\usepackage{cite}
\usepackage{cleveref}
\usepackage{color}
\usepackage{diagbox}
\usepackage{epsfig,latexsym}
\usepackage{epstopdf}
\usepackage{graphicx}
\usepackage{fancyhdr}
\usepackage{float}
\usepackage{flushend}
\usepackage{indentfirst}
\usepackage{lastpage}
\usepackage{makecell}
\usepackage{mathtools}
\usepackage{multirow}
\usepackage{psfrag}
\usepackage{setspace}
\usepackage{stfloats}
\usepackage{subfloat}
\usepackage{tikz}
\usepackage{times}

\theoremstyle{remark}

\allowdisplaybreaks[4]

\begin{document}

\title{Reconfigurable Intelligent Surface Aided Amplitude- and Phase-Modulated Downlink Transmission}

\author{Qingchao Li, Mohammed El-Hajjar, \textit{Senior Member, IEEE}, Ibrahim Hemadeh, \textit{Member, IEEE}, Arman Shojaeifard, \textit{Senior Member, IEEE}, Alain A. M. Mourad, Lajos Hanzo, \textit{Life Fellow, IEEE}

\thanks{Qingchao Li, Mohammed El-Hajjar and Lajos Hanzo are with the Electronics and Computer Science, University of Southampton, Southampton SO17 1BJ, U.K. (e-mail: Qingchao.Li@soton.ac.uk; meh@ecs.soton.ac.uk; lh@ecs.soton.ac.uk).

Ibrahim Hemadeh, Arman Shojaeifard and Alain A. M. Mourad are with InterDigital, London EC2A 3QR, U.K. (e-mail: Ibrahim.Hemadeh@interdigital.com; Arman.Shojaeifard@interdigital.com; Alain.Mourad@interdigital.com).}}

\maketitle

\begin{abstract}
New reconfigurable intelligent surface (RIS) based amplitude and phase modulation schemes are proposed as an evolution how the phase-only modulation schemes available in the literature. Explicitly, both the amplitude-phase shift keying (A-PSK) and quadrature amplitude-phase shift keying (QA-PSK) are conceived, where the RIS is assumed to be part of a transmitter to deliver information to the multi-antenna aided downlink receiver. In the proposed design, the RIS is partitioned into multiple blocks, and the information bits are conveyed by controlling both the ON-OFF state and the phase shift of the RIS elements in each block. Since the propagation paths spanning from each RIS block to the receiver can be coherently combined as a benefit of appropriately configuring the phase of the RIS elements, the received signal constellations can be designed by controlling both the ON-OFF pattern of the RIS blocks as well as the phase shift of the RIS elements. Both the theoretical analysis and the simulation results show that our proposed RIS-aided modulation schemes outperform the state-of-the-art RIS-based PSK modulation both in terms of its discrete-input-continuous-output memoryless channel (DCMC) capacity and its symbol error probability, especially in the high signal-to-noise-ratio (SNR) region, when considering realistic finite resolution RIS phase shifts.
\end{abstract}
\begin{IEEEkeywords}
Reconfigurable intelligent surfaces (RIS), amplitude-phase modulation, channel capacity, symbol error probability.
\end{IEEEkeywords}

\section{Introduction}
Reconfigurable intelligent surfaces (RIS) are capable of beneficially reconfiguring the wireless environment by deploying a large number of passive reflecting elements for suitably adjusting the phase shift and even potentially the amplitude of the impinging signals \cite{hou2021mimo,li2022reconfigurable,li2022reconfigurable_iot}. Furthermore, RISs may also act as a transmitter relying on a single RF chain, where the information is conveyed by appropriately configuring the reflection coefficients of the passive RIS elements. This has promising applications in wireless communications as a benefit of its extremely low hardware complexity compared to conventional MIMO systems \cite{basar2019wireless,basar2019transmission,khaleel2020reconfigurable,
tang2020mimo,basar2020reconfigurable,yuan2021receive,li2021single}.

For the sake of reducing the RIS configuration complexity, most published work considers the amplitudes of RIS elements to be fixed and the signals are only manipulated by controlling the RIS phase shifts \cite{basar2019wireless,basar2019transmission,khaleel2020reconfigurable,
tang2020mimo,basar2020reconfigurable,yuan2021receive,li2021single}. Therefore, in this case, phase shift keying (PSK) modulation can be readily realized using RISs, since the PSK signals have a constant envelope. Specifically, the phase shift of each RIS element is adjusted by taking into account the corresponding channel phase of the link spanning from the RIS to the receiver for maximizing the channel gain, where additionally an $M$-level phase shift may be imposed on the signals reflected from all RIS elements for creating a virtual $M$-ary PSK signal constellation \cite{basar2019wireless}. In \cite{basar2019transmission}, Basar \textit{et al.} proposed an amalgamated blind access point and RIS modulation scheme capable of operating without channel state information (CSI), where a binary phase shift of 0 and $\pi$ is imposed on all RIS elements to create a binary phase shift keying constellation. However, this was attained at the cost of a certain performance loss. In \cite{khaleel2020reconfigurable}, the RIS was partitioned into two blocks, and the classic Alamouti scheme was employed based on configuring the phase shift of the RIS elements, with the information mapped to the virtual $M$-PSK symbols.

Further improvements can achieved by exploiting that in the quadrature amplitude modulation (QAM), the amplitudes of the signals convey extra information, but it is not intuitive at all how we can intrinsically amalgamate QAM with a RIS-based transmitter via the above methods relying on the constant envelope constraint. In \cite{tang2020mimo}, Tang \textit{et al.} conceived a high-order QAM constellation based on independently controlling the amplitude and phase shift of each RIS element by introducing a non-linear modulation technique under the constraint of a constant envelope, which was however quite complex.

In \cite{basar2020reconfigurable}, Basar constructed a RIS-based single-RF transmitter relying either on space shift keying or on spatial modulation (SM). Explicitly, the signals radiated from the RF-chain are unmodulated and information is only conveyed to the specific receiver antenna (RA). The phase shift of each RIS element is configured to design the passive beamforming from the RIS to the selected RA. To further increase the throughput, Yuan \textit{et al.} \cite{yuan2021receive} proposed a RIS-aided receiver-side quadrature reflecting modulation scheme, where the RIS is partitioned into two halves associated with the in-phase and quadrature components. Then the information is conveyed via each half of the RIS to form a beam focussed on a specific antenna at the receiver. However, the spatial modulation philosophy was applied at the user equipment side, which increased their receiver complexity.

In our context, the RIS is deployed as a transmitter, and we propose a pair of new RIS-based amplitude-phase modulation schemes, namely the amplitude-phase shift keying (A-PSK) and quadrature amplitude-phase shift keying (QA-PSK). Explicitly, our contributions are as follows:

\begin{itemize}
  \item We partition the RIS into multiple blocks, where the information is conveyed based on both the ON-OFF state and on the phase shift of each block, which is similar to the concept of the SM for MIMO systems in \cite{di2013spatial}. Furthermore, since the phase of the RIS elements in each block can be beneficially configured for coherently combining the fading channels, the received signal constellation can be conveniently controlled, which is different from conventional SM, where the received signal constellation cannot be controlled owing to the random fast fading. Furthermore, the maximum likelihood (ML) detector is derived for our proposed schemes.
  \item Both the discrete-input-continuous-output memoryless channel (DCMC) capacity and the symbol error probability (SEP) of our proposed schemes are derived. Our simulation results reveal that our arrangement outperforms the state-of-the-art (SoA) RIS-based PSK, especially in high rate transmission and for realistic finite RIS phase shift resolution.
\end{itemize}

\textit{Notations:} $(\cdot)^{\text{T}}$ and  $(\cdot)^{\text{H}}$ represent the transpose and Hermitian transpose operation, respectively, $\mathbb{C}^{m\times n}$ denotes the space of $m\times n$ complex-valued matrices, $\mathbf{I}_n$ represents the $n\times n$ identity matrix, $\mathbf{0}_n$ and $\mathbf{1}_n$ are the $n\times1$ vectors with all elements being 0 and 1, respectively, $\mathcal{R}(\mathbf{a})$ and $\mathcal{I}(\mathbf{a})$ represent the real and imaginary parts of the complex vector $\mathbf{a}$, respectively, $f_X(x)$ is the probability density function (PDF) of a random variables $X$, a complex Gaussian random vector with mean $\mathbf{a}$ and covariance matrix $\mathbf{\Sigma}$ is denoted as $\mathcal{CN}(\mathbf{a},\mathbf{\Sigma})$, $\mathbb{E}(X)$ represents the mean of the random variable $X$.

\section{System Model}\label{System_Model}
\begin{figure}[!t]
    \centering
    \includegraphics[width=2.9in]{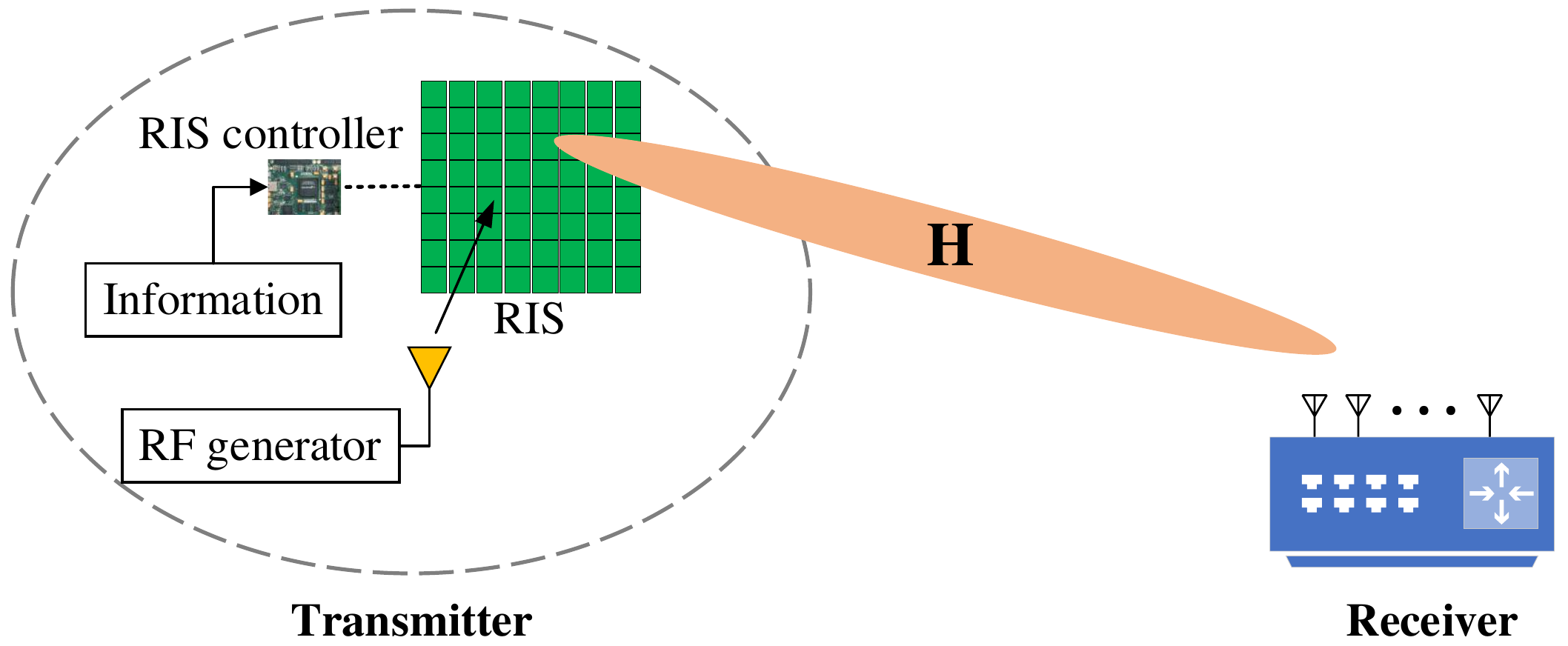}
    \caption{System model of RIS-based single RF-chain transmitter.}\label{Fig_1}
\end{figure}

As in \cite{basar2019wireless,basar2019transmission,khaleel2020reconfigurable,
tang2020mimo,basar2020reconfigurable,yuan2021receive,li2021single}, the RIS is configured as the low-complexity transmitter shown in Fig. \ref{Fig_1}, where a single RF chain generates the unmodulated carrier of wavelength $\lambda$ impinging on the $N$-element passive RIS. The RIS controller adjusts the phase shift of the RIS elements according to the baseband information and the CSI, where the information is conveyed by the specific RIS phase pattern configuration. The carrier wave impinging on the RIS is then reflected to the $K$-antenna single-user receiver. Since the transmitter RF generator of Fig. \ref{Fig_1} is close to the RIS, the RIS can be viewed as part of the transmitter, and thus the fading effects between the
RF generator and the RIS can be ignored \cite{basar2019wireless,basar2019transmission,
khaleel2020reconfigurable,tang2020mimo,basar2020reconfigurable,yuan2021receive,li2021single}. We denote the channel between the RIS and the receiver as $\mathbf{H}\in\mathbb{C}^{K\times N}$, and $\mathbf{H}=[\mathbf{h}_1^{\mathrm{H}},\mathbf{h}_2^{\mathrm{H}},\cdots,
\mathbf{h}_K^{\mathrm{H}}]^{\mathrm{H}}$, where $\mathbf{h}_k\in\mathbb{C}^{1\times N}$ represents the specific link between the $N$-element RIS and the $k$th antenna at the receiver. We assume that all the links are independent and experience flat Rician fading \cite{yuan2021receive}. Thus, $\mathbf{h}_k$ can be expressed as
\begin{align}\label{channel_model_1}
    \mathbf{h}_k=\sqrt{\frac{\kappa}{1+\kappa}}\overline{\mathbf{h}}_k+\sqrt{\frac{1}{1+\kappa}}
    \widetilde{\mathbf{h}}_k,
\end{align}
where $\kappa$ is the Rician factor and $\overline{\mathbf{h}}_k$ denotes the line-of-sight (LoS) component, satisfying $|\overline{\mathbf{h}}_k|=\mathbf{1}_N$, while $\widetilde{\mathbf{h}}_k$ denotes the non-line-of-sight (NLoS) component obeying $\widetilde{\mathbf{h}}_k\sim\mathcal{CN}(\mathbf{0}_N,\mathbf{I}_N)$. We also assume that instantaneous CSI can be attained at the transmitter, which may be estimated as in \cite{lin2020adaptive} for example. The receiver combining vector $\mathbf{w}\in\mathbb{C}^{1\times K}$ of the user relies on statistical CSI, namely on the angle of arrival (AoA) $\phi$ at the receiver, as follows
\begin{align}\label{channel_model_2}
    \mathbf{w}=[1,\mathrm{e}^{j\frac{2\pi}{\lambda}d\sin\phi},\cdots,\mathrm{e}^{j\frac{2\pi}
    {\lambda}d(K-1)\sin\phi}],
\end{align}
where $d$ is the distance between the adjacent RAs. Therefore, the equivalent channel vector of the link is given by
\begin{align}\label{channel_model_3}
    \mathbf{g}=\mathbf{w}\mathbf{H}=\sqrt{\frac{\kappa}{1+\kappa}}K\overline{\mathbf{h}}_1
    +\sqrt{\frac{1}{1+\kappa}}\sum_{k=1}^{K}\widetilde{\mathbf{h}}_ke^{j\frac{2\pi}{\lambda}
    d(k-1)\sin\phi}.
\end{align}
Since the links $\widetilde{\mathbf{h}}_k$ ($k=1,\cdots,K$) in (\ref{channel_model_3}) are independently and identically distributed obeying $\mathcal{CN}(\mathbf{0}_N,\mathbf{I}_N)$, the distribution of $\sum_{k=1}^{K}\widetilde{\mathbf{h}}_k\mathrm{e}^{j\frac{2\pi}{\lambda}d(k-1)\sin\phi}$ is given by $\mathcal{CN}(\mathbf{0}_N,K\mathbf{I}_N)$. We denote the LoS component of $\mathbf{g}$ as $K\overline{\mathbf{h}}_1$, and the NLoS component of $\mathbf{g}$ as $\sum_{k=1}^{K}\widetilde{\mathbf{h}}_k\mathrm{e}^{j\frac{2\pi}{\lambda}d(k-1)\sin\phi}$, which follows $\mathcal{CN}(\mathbf{0}_N,K\mathbf{I}_N)$. Therefore, the $K$-antenna receiver relying on the statistical CSI constituted by the received SNR $\rho$ and experiencing the Rician factor $\kappa$ may be declared equivalent to a single-antenna receiver having the received SNR $\rho'=(\frac{\kappa}{1+\kappa}K+\frac{1}{1+\kappa})\rho$ and the Rician factor $\kappa'=K\kappa$.

We denote the phase shift of the RIS elements as $\mathbf{\Theta}=[\theta_1,\cdots,\theta_N]^{\text{T}}$, where we assume that the phase shift of each RIS element has $B$-bit resolutions, i.e. the phase shift of each RIS element belongs to the set $\{0,\frac{2\pi}{2^B},\cdots,(2^B-1)\cdot\frac{2\pi}{2^B}\}$ \cite{tang2020mimo}.

In the following, firstly the SoA RIS-based modulation is presented, and then we propose a pair of new RIS-based amplitude-phase modulation schemes.

\subsection{State-of-the-art RIS-based Modulation}
\begin{figure*}[!t]
    \begin{minipage}[t]{0.33\linewidth}\centering
        \includegraphics[width=2.3\textwidth]{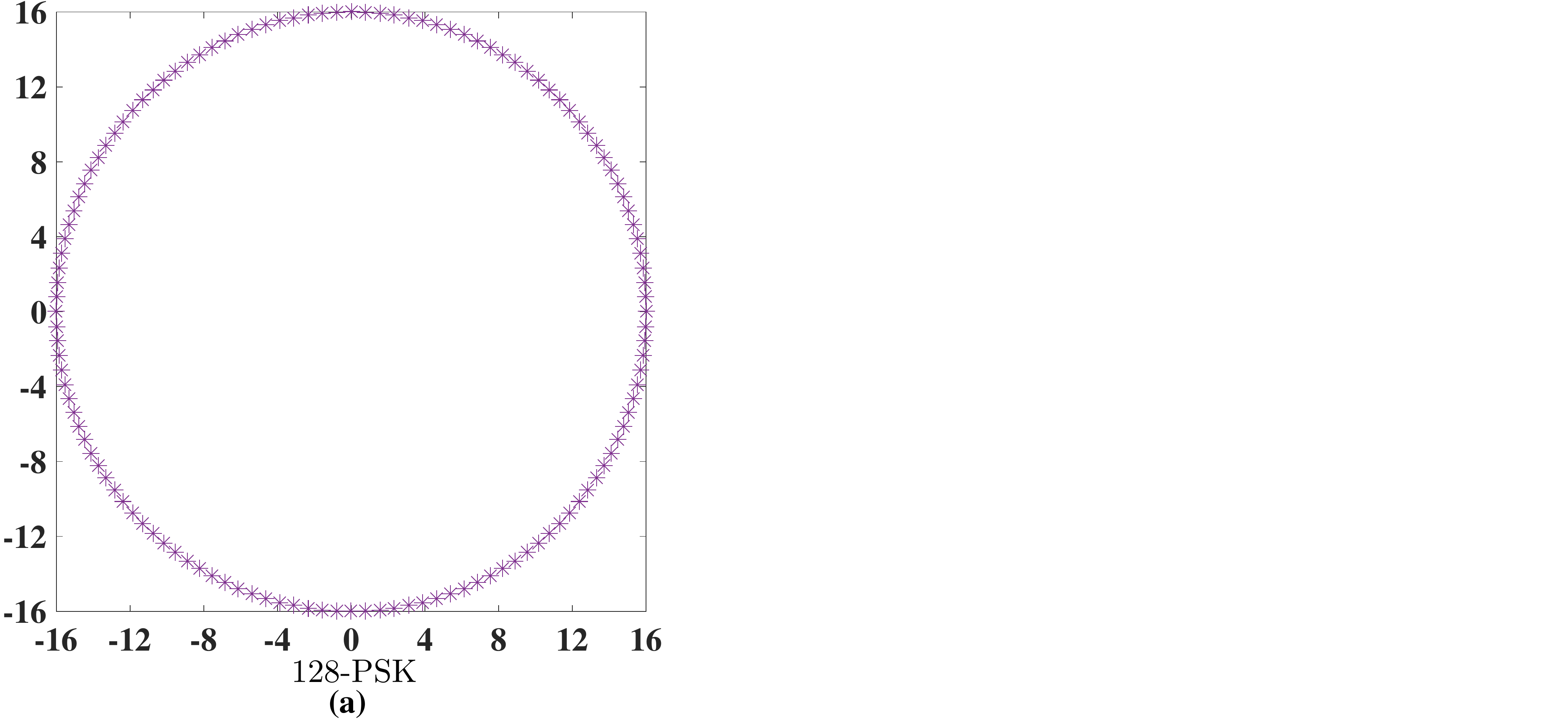}
    \end{minipage}
    \begin{minipage}[t]{0.33\linewidth}\centering
        \includegraphics[width=2.3\textwidth]{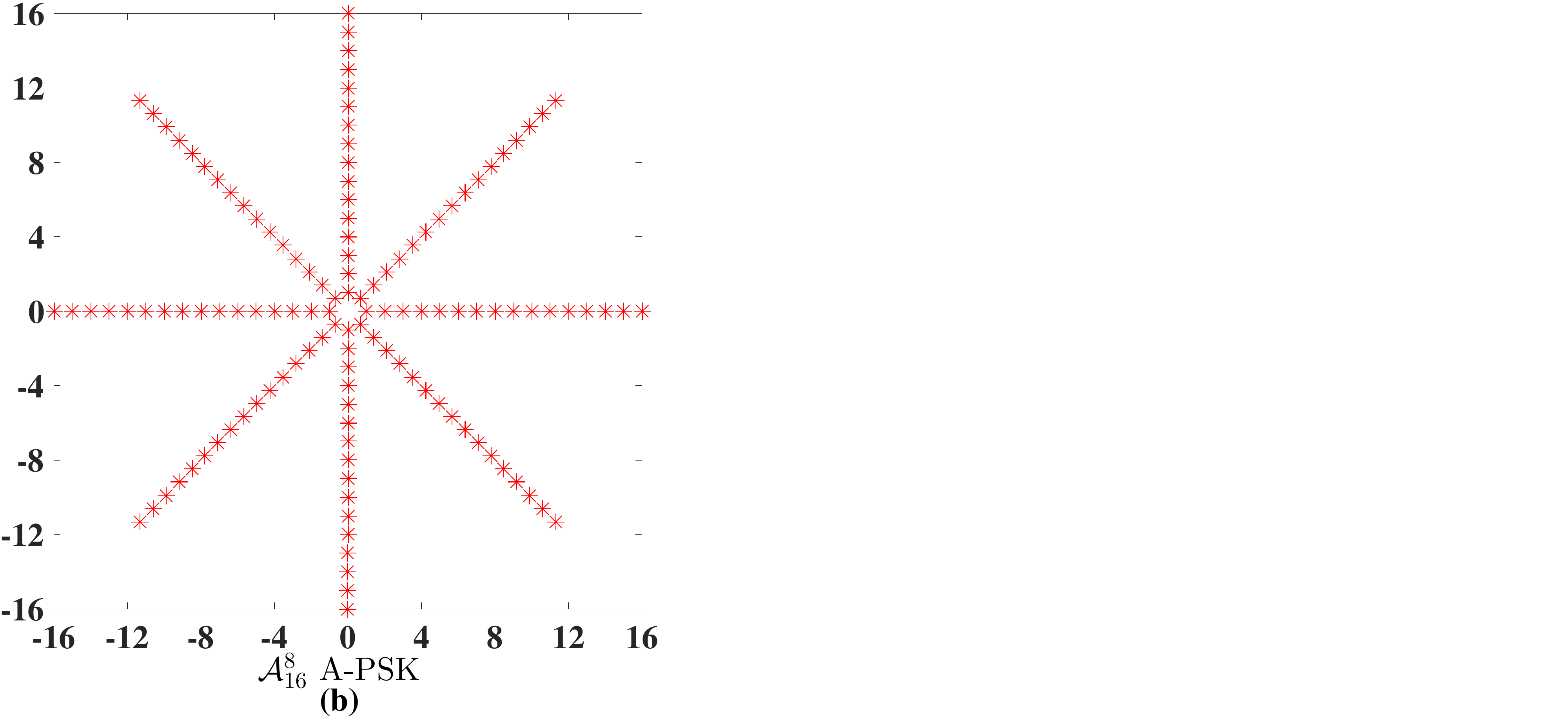}
    \end{minipage}
    \begin{minipage}[t]{0.33\linewidth}\centering
        \includegraphics[width=2.3\textwidth]{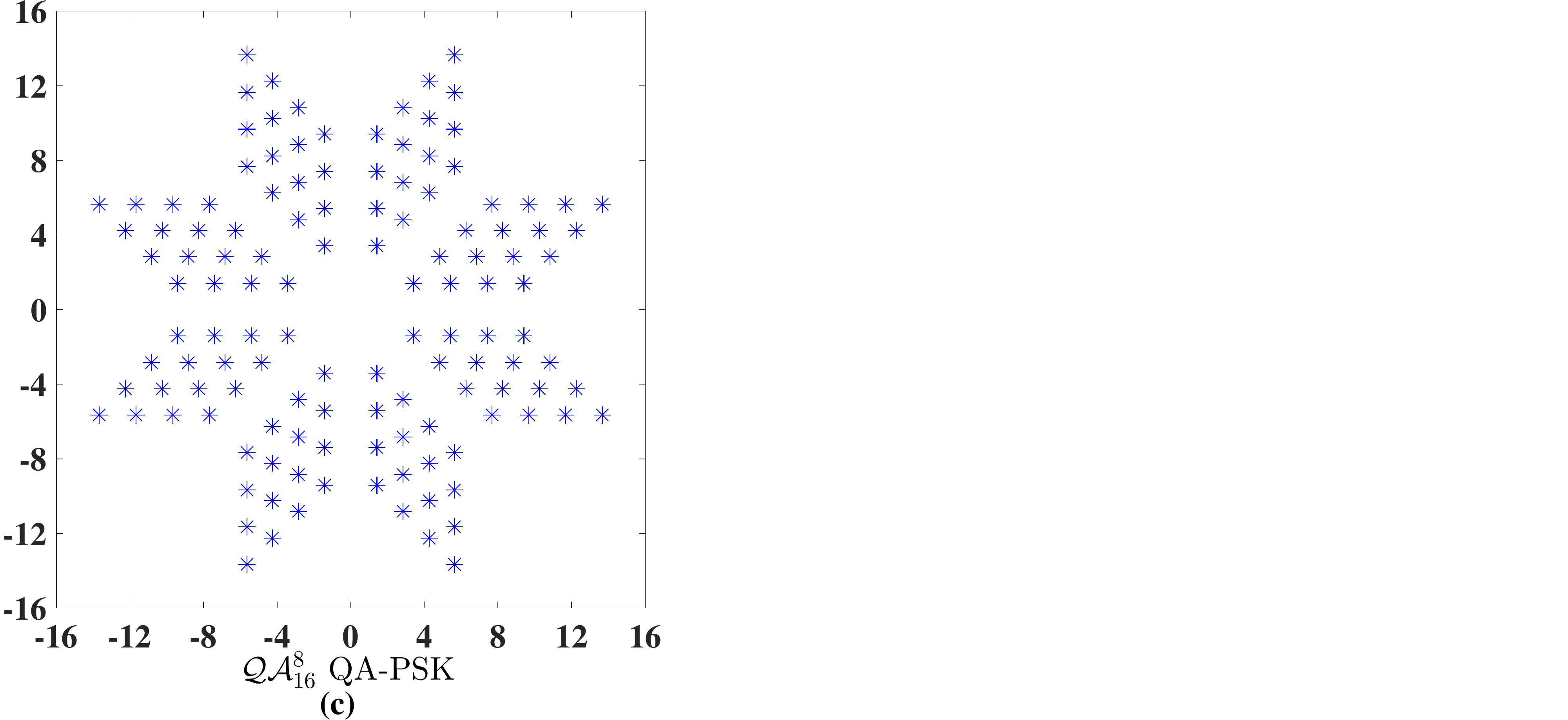}
    \end{minipage}
    \caption{The statistical CSI-based received signal constellation in (a) the SoA RIS-based 128-PSK, (b) the proposed RIS-based $\mathcal{A}_{16}^8$ A-PSK and (c) $\mathcal{Q}_{16}^8$ QA-PSK.}\label{Fig_2}
\end{figure*}

In the SoA RIS-based $M$-PSK modulation \cite{basar2019wireless,basar2019transmission,khaleel2020reconfigurable}, the channel fading experienced by all RIS elements is coherently combined for detecting the transmitted information symbol $m$ ($m=0,1,\cdots,M-1$), where the phase shift of the RIS elements is designed as
\begin{align}\label{system_model_PSK_1}
    \mathbf{\Theta}=\underset{^B}{\angle}(\mathrm{e}^{j\frac{2\pi m}{M}}\mathbf{g}^{\text{H}}),
\end{align}
with $\underset{^B}{\angle}(\cdot)$ representing the phase calculation using $B$-bit quantization. Thus, the signal set of $M$-PSK modulation is given by:
\begin{align}\label{system_model_PSK_2}
    \mathbb{S}_{M\text{-PSK}}&=\{\mathbf{g}\cdot\underset{^B}{\angle}
    (\mathrm{e}^{j\frac{2\pi m}{M}}\mathbf{g}^{\text{H}})|m=0,1,\cdots,M-1\}.
\end{align}
From (\ref{system_model_PSK_2}), we can find that when $M\leq2^B$, the received signals have a unique envelope, and $\mathbb{S}_{M\text{-PSK}}$ can be simplified as $\{\mathrm{e}^{j\frac{2\pi m}{M}}X|m=0,\cdots,M-1\}$, where $X=\mathbf{g}\cdot\underset{^B}{\angle}(\mathbf{g}^{\text{H}})$ is the constant envelope of the received signals. However, when $M>2^B$, the envelope of the received signals in $\mathbb{S}_{M\text{-PSK}}$ is not necessarily unique due to the $B$-bit phase-quantization. Fig. \ref{Fig_2} (a) shows an example of the statistical CSI-based received signal constellation of 128-PSK modulation.

\subsection{Proposed RIS-based Modulation}
The SoA RIS-based modulation relies on pure phase shift control, while higher order modulation schemes can be realized by conveying the information both on the amplitude and the phase shift of the modulated signals. However, the RIS requires the employment of active reflection type amplifiers in order to control the amplitude of each RIS element \cite{zhang2021active}, which requires high hardware cost and complexity. In this section we
present our proposed designs capable of both amplitude and phase modulation based on controlling the ON-OFF state and the phase shift of RIS elements without requiring any additional hardware components.

\subsubsection{RIS-based Amplitude-Phase Shift Keying Modulation}
Inspired by the conventional amplitude-phase shift keying (A-PSK), where the information is conveyed both by the amplitude and the phase shift of modulated signals, we propose a RIS-based A-PSK scheme. Firstly, we partition the RIS into $\frac{M}{V}$ blocks $\mathcal{B}_1,\cdots,\mathcal{B}_{\frac{M}{V}}$, each of which contains $\frac{NV}{M}$ elements. We denote the channel spanning from each block to the receiver as $\mathbf{g}_1,\cdots,\mathbf{g}_{\frac{M}{V}}$, respectively. To realize $M$-ary modulation by the $N$-element RIS, we partition the $M$-ary information into two parts, i.e. $\frac{M}{V}$-ary information, denoted as $l$ ($l=1,\cdots,\frac{M}{V}$), conveyed by the ON-OFF state of the RIS blocks and $V$-ary information, denoted as $v$ ($v=0,\cdots,V-1$), and conveyed by the phase shift of the ON-state RIS blocks. Specifically, in each information transmission slot, the blocks $\mathcal{B}_1,\cdots,\mathcal{B}_{l}$ are turned on, i.e. the amplitudes of the RIS elements in these blocks aer set to 1, while the blocks $\mathcal{B}_{l+1},\cdots,\mathcal{B}_{\frac{M}{V}}$ are turned off with the amplitudes of the RIS elements in these blocks set to 0. To avoid any RIS phase matching problem, $V$ is chosen as a divisor of $2^B$, hence $\log_2V$ is an integer not larger than $B$. Furthermore, in all ON-state RIS blocks, the $V$-PSK modulation scheme of (\ref{system_model_PSK_1}) is employed. Therefore, the RIS phase shift of the ON-state blocks $\mathcal{B}_1,\cdots,\mathcal{B}_{l}$, denoted as $\mathbf{\Theta}_1,\cdots,\mathbf{\Theta}_l$, is given by
\begin{align}\label{star_QAM_1}
    \mathbf{\Theta}_{(1:l)}^{\text{T}}=\underset{^B}{\angle}\big(\mathrm{e}^{j\frac{2\pi v}{V}}\mathbf{g}_{(1:l)}^{\text{H}}\big)=\mathrm{e}^{j\frac{2\pi v}{V}}\underset{^B}{\angle}\big(\mathbf{g}_{(1:l)}^{\text{H}}\big),
\end{align}
where $\mathbf{\Theta}_{(1:l)}=[\mathbf{\Theta}_1,\cdots,\mathbf{\Theta}_l]$, and $\mathbf{g}_{(1:l)}=[\mathbf{g}_1,\cdots,\mathbf{g}_l]$. Thus, an $M$-ary information symbol can be transmitted per each channel use upon appropriately controlling $l$ and $v$ in (\ref{star_QAM_1}). Fig. \ref{Fig_2} (b) shows an example of the statistical CSI based constellation of the received signals using this modulation scheme, where we have $M=128$ and $V=8$. Again, the information is jointly conveyed both by the $\frac{M}{V}$-level amplitude and the $V$-level phase shift. Hence we term this scheme as ($\frac{M}{V}$,$V$) A-PSK modulation, denoted as $\mathcal{A}_{\frac{M}{V}}^V$ A-PSK, where the set of received signals is given by:
\begin{align}\label{star_QAM_2}
    \mathbb{S}_{\mathcal{A}_{\frac{M}{V}}^{V}}&=\Big\{\mathrm{e}^{j\frac{2\pi v}{V}}\sum_{l'=1}^{l}X_{l'}\Big|v=\!0,\cdots,V-1;l=1,\cdots,\frac{M}{V}\Big\},
\end{align}
and $X_{l'}=\mathbf{g}_{l'}\mathbf{\Theta}_{l'}$ is the channel gain of the block $\mathcal{B}_{l'}$.

\subsubsection{RIS-based Quadrature Amplitude-Phase Shift Keying Modulation}
Inspired by the classic QAM scheme, we improve the above RIS-based A-PSK modulation as follows. Firstly, we partition the $N$-element RIS into two branches, namely the in-phase (I-) and quadrature (Q-) branch, with each containing $\frac{N}{2}$ RIS elements. Explicitly, in each branch, the RIS is divided into $\sqrt{\frac{M}{V}}$ blocks, each of which contains $\frac{N}{2}\sqrt{\frac{V}{M}}$ elements, and we denote these blocks as $\mathcal{B}_1^{\text{(I)}},\cdots,\mathcal{B}_{\sqrt{\frac{M}{V}}}^{\text{(I)}}$
and $\mathcal{B}_1^{\text{(Q)}},\cdots,\mathcal{B}_{\sqrt{\frac{M}{V}}}^{\text{(Q)}}$
in the I-branch and Q-branch, respectively. Furthermore, we denote these channels as
$\mathbf{g}_1^{\text{(I)}},\cdots,\mathbf{g}_{\sqrt{\frac{M}{V}}}^{\text{(I)}}$
and $\mathbf{g}_1^{\text{(Q)}},\cdots,\mathbf{g}_{\sqrt{\frac{M}{V}}}^{\text{(Q)}}$,
in the I-branch and Q-branch, respectively. To realize $M$-ary modulation, we partition the $M$-ary information into two parts. In the first part, $\frac{M}{V}$-ary information is conveyed, denoted as the pair $(l_1,l_2)$ ($l_1,l_2=1,\cdots,\sqrt{\frac{M}{V}}$), which is carried by the ON-OFF state of the RIS blocks. The second part conveys $V$-ary information, denoted as $v$ ($v=0,\cdots,V-1$), which is
represented by the phase shift of the ON-state RIS blocks. Specifically, in each information transmission slot, the blocks $\mathcal{B}_1^{\text{(I)}},\cdots,\mathcal{B}_{l_1}^{\text{(I)}}$ in the I-branch and the blocks $\mathcal{B}_1^{\text{(Q)}},\cdots,\mathcal{B}_{l_2}^{\text{(Q)}}$ in the Q-branch are turned on, while all other blocks are turned off. Again, to avoid RIS phase matching problems, $V$ is chosen as a divisor of $2^B$, with $\log_2V$ being an integer not larger than $B$. Additionally, in the ON-state blocks of each branch, $V$-PSK modulation is employed. Note that, to form a two-dimensional amplitude, a phase shift of $e^{j\frac{2\pi}{V}}$ must be between the I-branch and the Q-branch. Therefore, the RIS phase shift of blocks $\mathcal{B}_1^{\text{(I)}},\cdots,\mathcal{B}_{l_1}^{\text{(I)}}$ and $\mathcal{B}_1^{\text{(Q)}},\cdots,\mathcal{B}_{l_2}^{\text{(Q)}}$ are denoted by $\mathbf{\Theta}_{1}^{\text{(I)}},\cdots,\mathbf{\Theta}_{l_1}^{\text{(I)}}$ and $\mathbf{\Theta}_{1}^{\text{(Q)}},\cdots,\mathbf{\Theta}_{l_2}^{\text{(Q)}}$, and they are given by
\begin{align}\label{diamond_QAM_1}
    \mathbf{\Theta}_{(1:l_1)}^{\text{(I)}\text{T}}=\underset{^B}{\angle}(\mathrm{e}^{j\frac{2\pi v}{V}}\mathbf{g}_{(1:l_1)}^{\text{(I)}\text{H}})=\mathrm{e}^{j\frac{2\pi v}{V}}\underset{^B}{\angle}(\mathbf{g}_{(1:l_1)}^{\text{(I)}\text{H}}),
\end{align}
\begin{align}\label{diamond_QAM_2}
    \mathbf{\Theta}_{(1:l_2)}^{\text{(Q)}\text{T}}=\mathrm{e}^{j\frac{2\pi}{V}}\underset{^B}
    {\angle}(\mathrm{e}^{j\frac{2\pi v}{V}}\mathbf{g}_{(1:l_2)}^{\text{(Q)}\text{H}})=
    \mathrm{e}^{j\frac{2\pi(v+1)}{V}}\underset{^B}{\angle}(\mathbf{g}_{(1:l_2)}^{\text{(Q)}\text{H}}),
\end{align}
where $\mathbf{\Theta}_{(1:l_1)}^{\text{(I)}}=[\mathbf{\Theta}_1^{\text{(I)}},\cdots,
\mathbf{\Theta}_{l_1}^{\text{(I)}}]$,$\mathbf{\Theta}_{(1:l_2)}^{\text{(Q)}}=
[\mathbf{\Theta}_1^{\text{(Q)}},\cdots,\mathbf{\Theta}_{l_2}^{\text{(Q)}}]$, $\mathbf{g}_{(1:l_1)}^{\text{(I)}}=[\mathbf{g}_1^{\text{(I)}},\cdots,\mathbf{g}_{l_1}
^{\text{(I)}}]$, and $\mathbf{g}_{(1:l_2)}^{\text{(Q)}}=[\mathbf{g}_1^{\text{(Q)}},\cdots,
\mathbf{g}_{l_2}^{\text{(Q)}}]$. Thus, effectively an $M$-ary information symbol can be transmitted per each channel use by appropriately controlling $(l_1,l_2)$ and $v$ in (\ref{diamond_QAM_1}) and (\ref{diamond_QAM_2}). Fig. \ref{Fig_2} (c) shows an example of the statistical CSI-based received signal constellation for $M=128$ and $V=8$, relying on $\frac{M}{V}$-level two-dimensional amplitude and $V$-level phase shifts. Hence, we term this scheme as ($\frac{M}{V}$,$V$) quadrature amplitude-phase shift keying (QA-PSK), denoted as $\mathcal{Q}_{\frac{M}{V}}^V$ QA-PSK. Note that our QA-PSK modulation requires $B\geq2$ bits, where the set of received signals is given by
\begin{align}\label{diamond_QAM_3}
    \notag\mathbb{S}_{\mathcal{Q}_{\frac{M}{V}}^V}=&\Big\{\mathrm{e}^{j\frac{2\pi v}{V}}\big(\sum\nolimits_{l_1'=1}^{l_1}X_{l_1'}^{\text{(I)}}+\mathrm{e}^{j\frac{2\pi}{V}}
    \sum\nolimits_{l_2'=1}^{l_2}X_{l_2'}^{\text{(Q)}}\big)|v=0,\\
    &\quad \cdots,V-1;l_1,l_2=1,\cdots,\sqrt{\frac{M}{V}}\Big\},
\end{align}
with $X_{l_1'}^{\text{(I)}}=\mathbf{g}_{l_1'}^{\text{(I)}}\mathbf{\Theta}_{l_1'}^{\text{(I)}}$ and $X_{l_2'}^{\text{(Q)}}=\mathrm{e}^{-j\frac{2\pi}{V}}\mathbf{g}_{l_2'}^{\text{(Q)}}
\mathbf{\Theta}_{l_2'}^{\text{(Q)}}$ being the channel gain of block $\mathcal{B}_{l_1'}^{\text{(I)}}$ and block $\mathcal{B}_{l_2'}^{\text{(Q)}}$, respectively.

\subsection{Receiver Design}
The ML detection method is employed at the receiver to recover the information. We denote the received signal as $y$. Based on the ML criterion, the information recovered by the SoA RIS-based $M$-PSK, as well as by our proposed $\mathcal{A}_{\frac{M}{V}}^V$ A-PSK and $\mathcal{Q}_{\frac{M}{V}}^{V}$ QA-PSK are $\hat{s}=\min\limits_{\hat{m}\in\{0,\cdots,M-1\}}\|y-\sqrt{\rho'}\mathbf{g}\underset{^B}
{\angle}(\mathrm{e}^{j\frac{2\pi\hat{m}}{M}}\mathbf{g}^{\text{H}})\|$, $\hat{s}=\min\limits_{\hat{l}\in\{1,\cdots,\frac{M}{V}\},\hat{v}\in\{0,\cdots,V-1\}}
\|y-\sqrt{\rho'}e^{j\frac{2\pi\hat{v}}{V}}\sum_{l'=1}^{\hat{l}}X_{l'}\|$, and $\hat{s}=\min\limits_{\hat{l}_1,\hat{l}_2\in\{1,\cdots,\sqrt{\frac{M}{V}}\},\hat{v}\in\{0,\cdots,
V-1\}}\|y-\sqrt{\rho'}\mathrm{e}^{j\frac{2\pi\hat{v}}{V}}[(\sum_{l_1'=1}^{\hat{l}_1}X_{l_1'}
^{\text{(I)}}+\mathrm{e}^{j\frac{2\pi}{V}}\sum_{l_2'=1}^{\hat{l}_2}X_{l_2'}^{\text{(Q)}})]\|$, respectively.

\section{Theoretical Performance Analysis}\label{Performance Theoretical Analysis}
In this section, we derive the DCMC capacity and SEP expressions of our proposed RIS-based A-PSK and QA-PSK modulation schemes. We commence by first deriving the distribution of the channel gain for each RIS block of the proposed A-PSK and QA-PSK schemes.

In $\mathcal{A}_{\frac{M}{V}}^{V}$ A-PSK, the number of RIS elements in each block is $N_\mathcal{A}=\frac{NV}{M}$. The first moment of the channel gain $X_l$ is given by
\begin{align}\label{Theory_1}
    \mathbb{E}[X_l]=\mathbb{E}[\mathbf{g}_{l}\underset{^B}{\angle}(\mathbf{g}_{l}^{\text{H}})]
    =\mathbb{E}\Big[\sum\nolimits_{i=1}^{N_\mathcal{A}}\alpha_i\mathrm{e}^{j\psi_i}\Big],
\end{align}
where $\psi_i$ follows the uniform distribution within $(-\frac{\pi}{2^B},\frac{\pi}{2^B})$, and $\alpha_i$ is the amplitude of the path spanning from the $i$th element in block $\mathcal{B}_l$ to the receiver, which follows the Rician distribution with parameters of $\nu=\sqrt{\frac{\kappa'}{1+\kappa'}}$ and $\sigma=\sqrt{\frac{1}{2(1+\kappa')}}$, where the first moment and second moment are $\mathbb{E}[\alpha_i]=\sigma\sqrt{\frac{\pi}{2}}L_{\frac{1}{2}}(-\frac{\nu^2}{2\sigma^2})$ and $\mathbb{E}[\alpha_i^2]=2\sigma^2+\nu^2$, respectively. Therefore, (\ref{Theory_1}) can be expressed as
\begin{align}\label{Theory_2}
    \mathbb{E}[X_l]=\frac{2^B}{\pi}\sin\frac{\pi}{2^B}\sum_{i=1}^{N_\mathcal{A}}\mathbb{E}[\alpha_i]
    =N_\mathcal{A}\frac{2^B}{\pi}\sin\frac{\pi}{2^B}\frac{\sqrt{\pi}}{2}L_{\frac{1}{2}}(-\kappa'),
\end{align}
where $L_{\frac{1}{2}}(\cdot)$ is the Laguerre polynomial \cite{simon2005digital}. Then, the second moment of $X_l$ is given by
\begin{align}\label{Theory_3}
    \notag \mathbb{E}[X_l^2]=&\mathbb{E}
    \Big[\big(\sum\nolimits_{i=1}^{N_\mathcal{A}}\alpha_i\mathrm{e}^{j\psi_i}\big)^2\Big]\\
    \notag=&2\sum_{i_1=1}^{N_\mathcal{A}-1}\sum_{i_2=i_1+1}^{N_\mathcal{A}}
    \mathbb{E}[\alpha_{i_1}]\mathbb{E}[\alpha_{i_2}]
    \mathbb{E}[\cos(\psi_{i_1})]\mathbb{E}[\cos(\psi_{i_2})]+\\
    \notag&\sum_{i=1}^{N_\mathcal{A}}\mathbb{E}[\alpha_i^2]\mathbb{E}[(\cos\psi_i)^2]\\
    \notag=&N_\mathcal{A}(N_\mathcal{A}-1)\Big(\frac{2^B}{\pi}\sin\frac{\pi}{2^B}
    \frac{\sqrt{\pi}}{2}L_{\frac{1}{2}}\!(-\kappa')\Big)^2\\
    &+N_\mathcal{A}\cdot\frac{1+\frac{2^B}{2\pi}\sin\frac{2\pi}{2^B}}{2}.
\end{align}
According to (\ref{Theory_2}) and (\ref{Theory_3}), the channel gain $X_l$ can be approximated by a Gamma distribution having the PDF of
\begin{align}\label{Theory_5}
    f_{X_l}(x)=\frac{1}{\Gamma(k_{\mathcal{A}})\theta_{\mathcal{A}}^{k_{\mathcal{A}}}}
    x^{k_{\mathcal{A}}-1}\mathrm{e}^{-\frac{x}{\theta_{\mathcal{A}}}},
\end{align}
where the shape parameter $k_{\mathcal{A}}$ and the scale parameter $\theta_{\mathcal{A}}$ are
\begin{align}\label{Theory_6}
    k_{\mathcal{A}}=\frac{(\mathbb{E}[X_l])^2}{\mathbb{E}[X_l^2]-(\mathbb{E}[X_l])^2},\quad \theta_{\mathcal{A}}=\frac{\mathbb{E}[X_l^2]-(\mathbb{E}[X_l])^2}{\mathbb{E}[X_l]}.
\end{align}

In the $\mathcal{Q}_{\frac{M}{V}}^{V}$ QA-PSK, the number of RIS elements in each block, denoted as $N_\mathcal{Q}$, is $N_\mathcal{Q}=\frac{N}{2}\sqrt{\frac{V}{M}}$. The first moment and second moment of the channel gain $X_l^{\text{(I)}}$ and $X_l^{\text{(Q)}}$ can be similarly derived as in the case of A-PSK, upon simply replacing $N_{\mathcal{A}}$ by $N_{\mathcal{Q}}$ in (\ref{Theory_2}) and (\ref{Theory_3}). The channel gain $X_l^{\text{(I)}}$ and $X_l^{\text{(Q)}}$ can be approximated by a Gamma distribution having the PDF of ({\ref{Theory_5}}) upon simply replacing $k_{\mathcal{A}}$ and $\theta_{\mathcal{A}}$ by $k_{\mathcal{Q}}$ and $\theta_{\mathcal{Q}}$ in (\ref{Theory_6}), respectively, where the shape parameter $k_{\mathcal{Q}}$ and the scale parameter $\theta_{\mathcal{Q}}$ are similarly given upon replacing $\mathbb{E}[X_l]$ by $\mathbb{E}[X_l^{\text{(I)}}]$ and $\mathbb{E}[X_l^{\text{(Q)}}]$, as well as replacing $\mathbb{E}[(X_l)^2]$ by $\mathbb{E}[(X_l^{\text{(I)}})^2]$ and $\mathbb{E}[(X_l^{\text{(Q)}})^2]$ in (\ref{Theory_6}), respectively.

\begin{figure}[!t]\centering
    \includegraphics[width=1.9in]{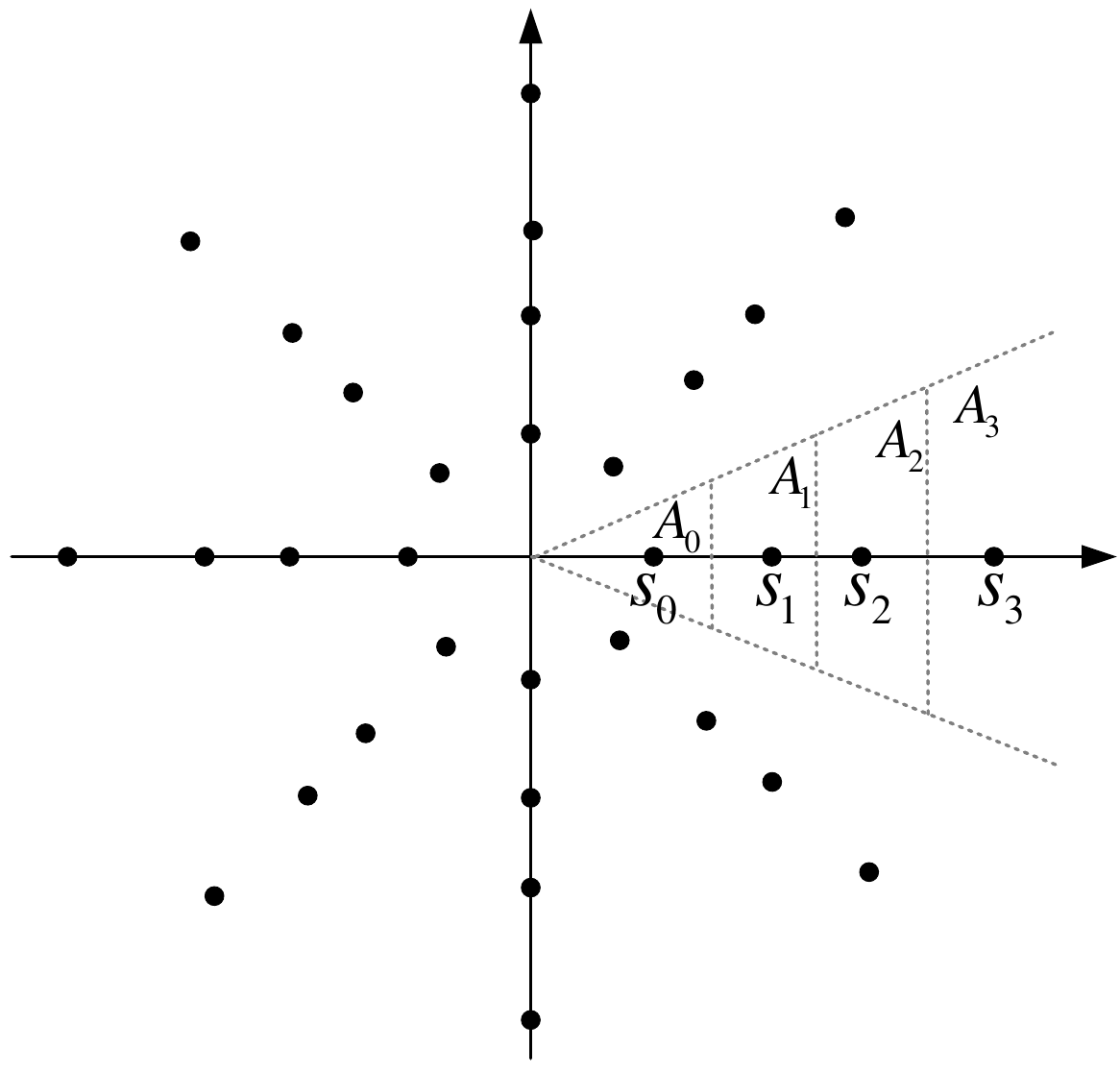}
    \caption{Illustration of ML detection for $\mathcal{A}_{4}^{8}$ A-PSK.}\label{Fig_3}
\end{figure}

\subsection{DCMC Capacity}
The DCMC capacity is given by \cite{el2008turbo}
\begin{align}\label{Channel_Capacity_DCMC_1}
    \notag R=&\mathbb{E}\Bigg[\log_2(M)-\frac{1}{M\pi}\sum_{m_1=1}^{M}\int_{-\infty}^{\infty}
    \int_{-\infty}^{\infty}\exp(-t_1^2-t_2^2)\cdot\\
    \notag&\log_2\Big(\sum_{m_2=1}^{M}\exp\Big(-2[t_1,t_2]
    \Big[\begin{array}{c}\sqrt{\rho'}\mathcal{R}(\mathbf{z}_{m_1}-\mathbf{z}_{m_2})\\
    \sqrt{\rho'}\mathcal{I}(\mathbf{z}_{m_1}-\mathbf{z}_{m_2})\\\end{array}\Big]-\\
    &\Big\|\Big[\begin{array}{c}\sqrt{\rho'}\mathcal{R}(\mathbf{z}_{m_1}-\mathbf{z}_{m_2})\\
    \sqrt{\rho'}\mathcal{I}(\mathbf{z}_{m_1}-\mathbf{z}_{m_2})\\\end{array}\Big]
    \Big\|^2\Big)\Big)\mathrm{d}t_1\mathrm{d}t_2\Bigg],
\end{align}
where $\mathbf{z}$ is an $N\times1$ vector with its elements being the set of received signals, which is given in (\ref{star_QAM_2}) and (\ref{diamond_QAM_3}) for $\mathcal{A}_{\frac{M}{V}}^{V}$ A-PSK and $\mathcal{Q}_{\frac{M}{V}}^{V}$ QA-PSK, respectively. Then (\ref{Channel_Capacity_DCMC_1}) can be expressed with the aid of the Gauss-Hermite Quadrature as \cite{stroud1966gaussian}
\begin{align}\label{Channel_Capacity_DCMC_2}
    \notag R=&\mathbb{E}\Bigg[\log_2(M)-\frac{1}{M\pi}\sum_{p_1=1}^{P}w_{p_1}\sum_{p_2=1}^{P}w_{p_2}
    f(t_{p_1},t_{p_2})\sum_{m_1=1}^{M}\\
    \notag&\log_2\Big(\sum_{m_2=1}^{M}\exp\Big(-2[t_1,t_2]
    \Big[\begin{array}{c}\sqrt{\rho'}\mathcal{R}(\mathbf{z}_{m_1}-\mathbf{z}_{m_2})\\
    \sqrt{\rho'}\mathcal{I}(\mathbf{z}_{m_1}-\mathbf{z}_{m_2})\\\end{array}\Big]\\
    &-\Big\|\Big[\begin{array}{c}\sqrt{\rho'}\mathcal{R}(\mathbf{z}_{m_1}-\mathbf{z}_{m_2})\\
    \sqrt{\rho'}\mathcal{I}(\mathbf{z}_{m_1}-\mathbf{z}_{m_2})\\\end{array}\Big]
    \Big\|^2\Big)\Big)\Bigg],
\end{align}
where $2P-1$ is the degree of precision, while the table of points $t_{p_1}$, $t_{p_2}$ and the weights $w_{p_1}$, $w_{p_2}$ are given in \cite{stroud1966gaussian}. Furthermore, $f(t_{p_1},t_{p_2})$ is given by
\begin{align}\label{Channel_Capacity_DCMC_3}
    \notag&f(t_{p_1},t_{p_2})\\
    \notag=&\sum_{m_1=1}^{M}\log_2\Bigg[\sum_{m_2=1}^{M}\exp\Big(-2[t_{p_1},t_{p_2}]
    \Big[\begin{array}{c}\sqrt{\rho'}\mathcal{R}(\mathbf{z}_{m_1}\!-\!\mathbf{z}_{m_2})
    \\\sqrt{\rho'}\mathcal{I}(\mathbf{z}_{m_1}\!-\!\mathbf{z}_{m_2})\\\end{array}\Big]\\
    &-\Big\|\Big[\begin{array}{c}\sqrt{\rho'}\mathcal{R}(\mathbf{z}_{m_1}\!-\!\mathbf{z}_{m_2})
    \\\sqrt{\rho'}\mathcal{I}(\mathbf{z}_{m_1}\!-\!\mathbf{z}_{m_2})\\\end{array}\Big]
    \Big\|^2\Big)\Bigg].
\end{align}
Since $f(t_{p_1},t_{p_2})$ is a concave function, the upper bound (UB) of the DCMC capacity, denoted as $R^{\text{(UB)}}$, is given by
\begin{align}\label{Channel_Capacity_DCMC_4}
    \notag R^{\text{(UB)}}=&\log_2(M)-\frac{1}{M\pi}\sum_{p_1=1}^{P}w_{p_1}\sum_{p_2=1}^{P}w_{p_2}
    f(t_{p_1},t_{p_2})\sum_{m_1=1}^{M}\\
    \notag&\ \log_2\Bigg[\sum_{m_2=1}^{M}\exp\Big(-2[t_1,t_2]
    \Big[\begin{array}{c}\sqrt{\rho'}\mathcal{R}(\mathbb{E}(\mathbf{z}_{m_1})
    -\mathbb{E}(\mathbf{z}_{m_2}))\\\sqrt{\rho'}\mathcal{I}
    (\mathbb{E}(\mathbf{z}_{m_1})-\mathbb{E}(\mathbf{z}_{m_2}))\\
    \end{array}\Big]\\
    &\ -\Big\|
    \Big[\begin{array}{c}\sqrt{\rho'}\mathcal{R}(\mathbb{E}(\mathbf{z}_{m_1})
    -\mathbb{E}(\mathbf{z}_{m_2}))\\\sqrt{\rho'}\mathcal{I}
    (\mathbb{E}(\mathbf{z}_{m_1})-\mathbb{E}(\mathbf{z}_{m_2}))\\
    \end{array}\Big]\Big\|^2\Big)\Bigg],
\end{align}
where the elements in $\mathbb{E}(\mathbf{z}_{m_1})$ and $\mathbb{E}(\mathbf{z}_{m_2})$ are from the set of $\{e^{j\frac{2\pi v}{V}}\sum_{l'=1}^{l}\mathbb{E}[X_{l'}]|v=0,\cdots,V-1;l=1,\cdots,\frac{M}{V}\!\}$ and $\{e^{j\frac{2\pi v}{V}}(\sum_{l_1'=1}^{l_1}\mathbb{E}[X_{l_1'}^{\text{(I)}}]
+e^{j\frac{2\pi}{V}}\sum_{l_2'=1}^{l_2}\mathbb{E}[X_{l_2'}^{\text{(Q)}}])| v=0,\cdots,V-1;l_1,l_2=1,\cdots,\sqrt{\frac{M}{V}}\}$ for ${\mathcal{A}_{\frac{M}{V}}^{V}}$ A-PSK and ${\mathcal{Q}_{\frac{M}{V}}^{V}}$ QA-PSK, respectively.

\subsection{Symbol Error Probability}
\subsubsection{Symbol Error Probability of A-PSK}
\begin{figure}[t]\centering
    \includegraphics[width=2.4in]{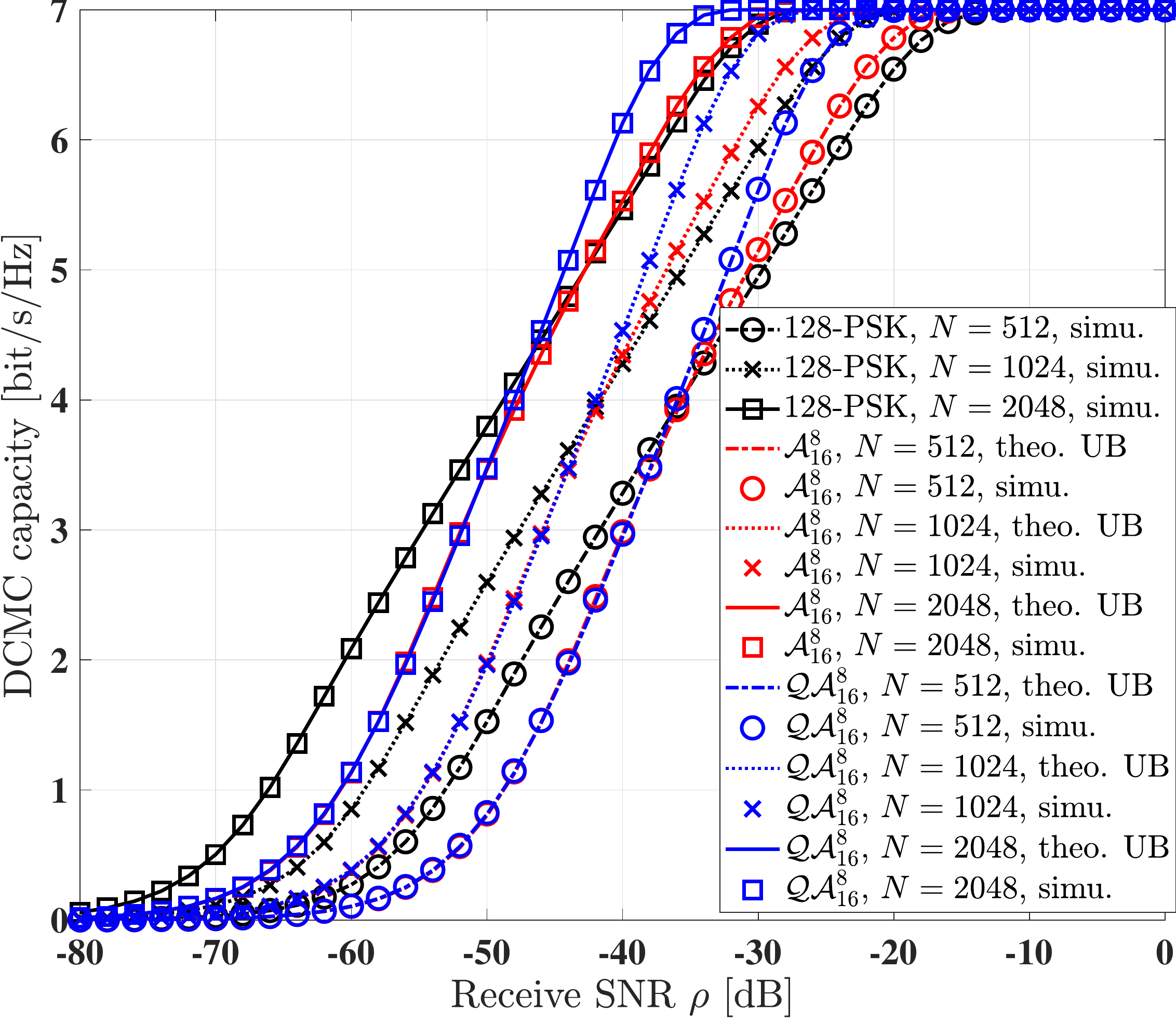}
    \caption{Comparison of DCMC capacity $R$ versus receive SNR $\rho$ for the SoA RIS-based 128-PSK, the proposed $\mathcal{A}_{16}^8$ A-PSK and $\mathcal{Q}_{16}^8$ QA-PSK with different number of RIS elements $N$, where the RIS phase shift resolution is $B=3$ bits.}\label{Fig_sepctral_efficiency}
\end{figure}
\begin{figure}[htb]
    \begin{minipage}[t]{1\linewidth}\centering
        \includegraphics[width=0.7245\textwidth]{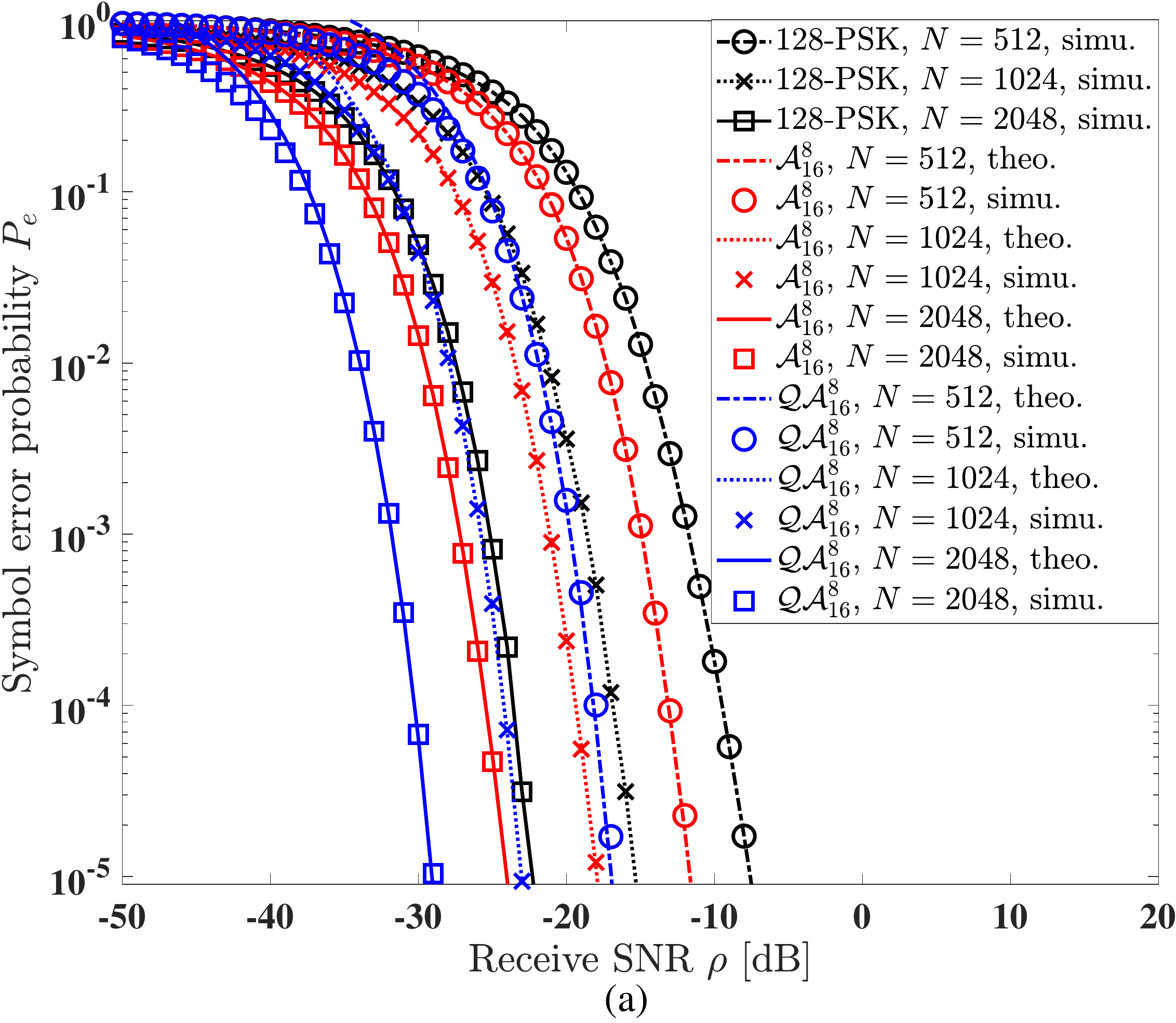}
    \end{minipage}
    \begin{minipage}[t]{1\linewidth}\centering
        \includegraphics[width=0.7245\textwidth]{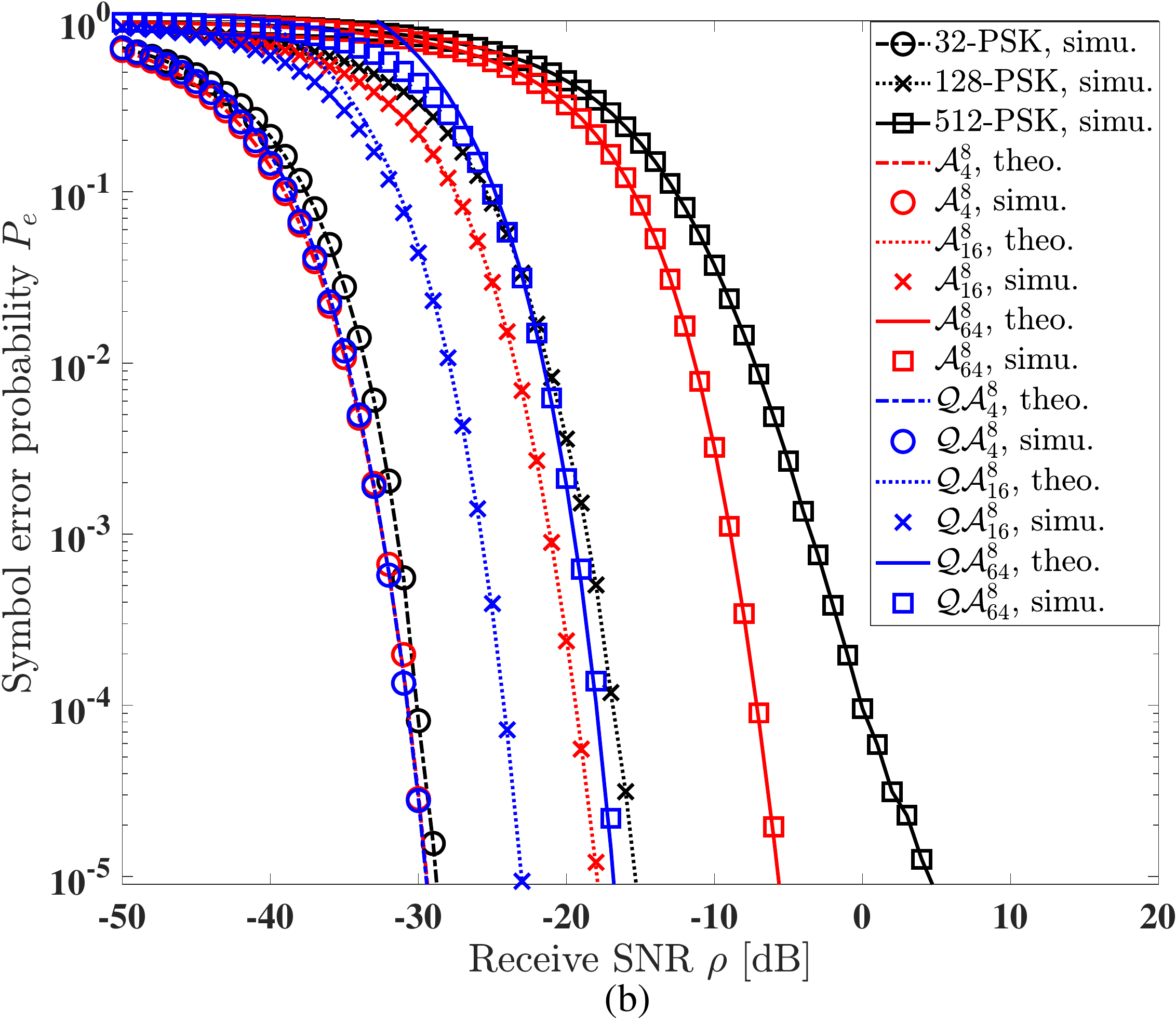}
    \end{minipage}
    \begin{minipage}[t]{1\linewidth}\centering
        \includegraphics[width=0.7245\textwidth]{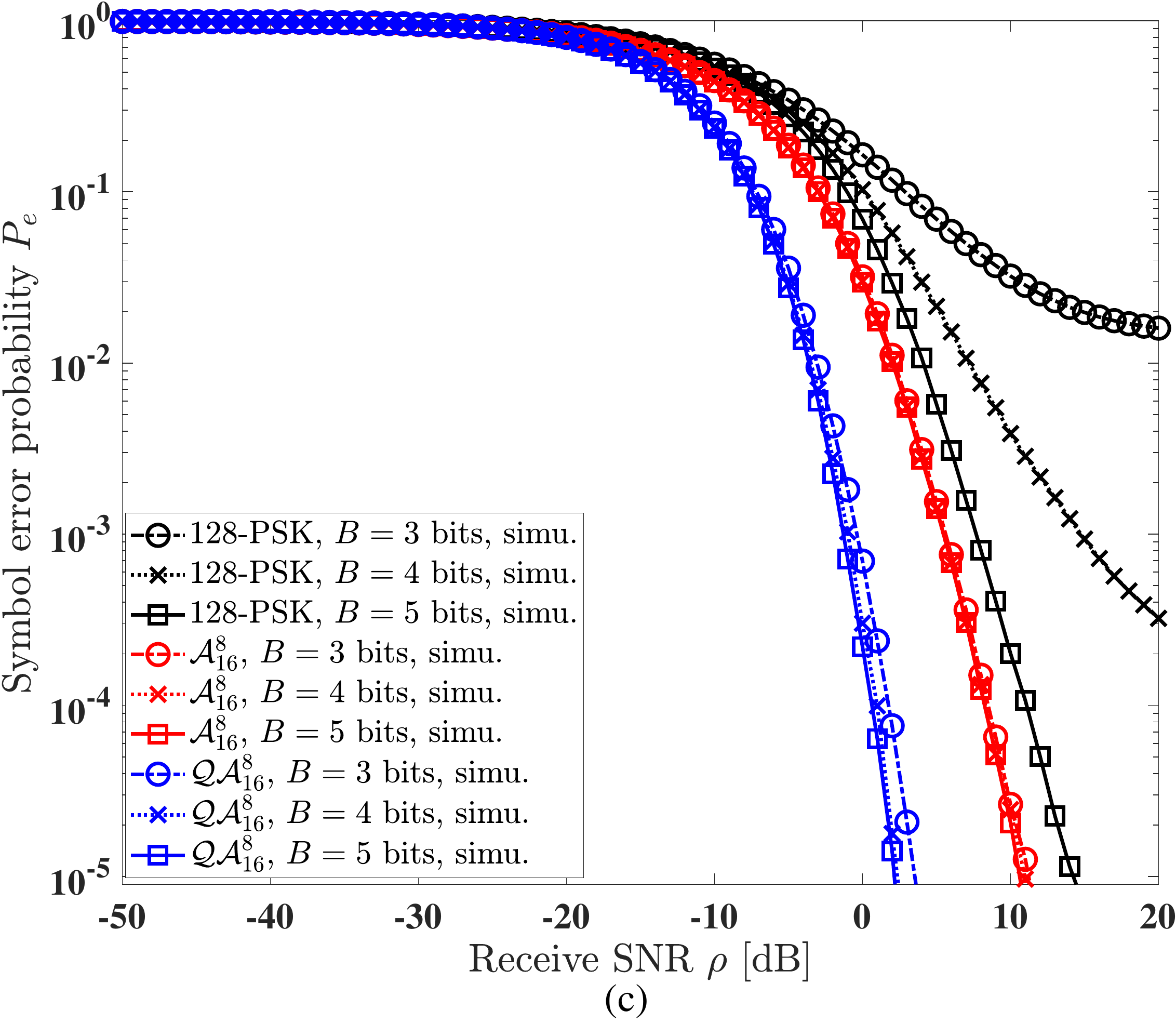}
    \end{minipage}
    \caption{Comparison of symbol error probability $P_e$ versus receive SNR $\rho$ for various modulation schemes: (a) with different number of RIS elements $N$; (b) with different transmission rate $R$; (c) with different RIS phase shift resolution $B$.}\label{Fig_simulation}
\end{figure}

In Fig. \ref{Fig_2} (b), we can find that when the ML detection method is employed, the symbol error probability of $\mathcal{S}_{\frac{M}{V}}^{V}$ A-PSK, denoted as $P_e^{(\mathcal{A})}$, is given by
\begin{align}\label{Pe_star_1}
    P_e^{(\mathcal{A})}=\frac{1}{\frac{M}{V}}\cdot \big(P_{e,1}^{(\mathcal{A})}+P_{e,2}^{(\mathcal{A})}+\cdots
    +P_{e,{\frac{M}{V}}}^{(\mathcal{A})}\big),
\end{align}
where $P_{e,l}$ represents the detection error probability, when the signal in the $l$th layer is transmitted, with $l=1,2,\cdots,\frac{M}{V}$.

Fig. \ref{Fig_3} illustrates the decision regions of ML detection for $\mathcal{A}_{4}^{8}$ A-PSK. For example, when the symbol $s_0$ in the 1st layer is transmitted, the receiver can correctly recover it, when the received signal is located in the triangular region $A_0$.

Therefore, $P_{e,1}^{(\mathcal{A})}$ is given by
\begin{align}\label{Pe_star_2}
    P_{e,1}^{(\mathcal{A})}=\frac{1}{2\pi}\sum_{k=0}^{2}\int_0^{\theta_k}
    \exp\Big[-\frac{\rho'b_k^2\sin^2\psi_k}{\sin^2\big(\theta+\psi_k\big)}\Big]\text{d}\theta,
\end{align}
where $b_0=b_2=\sqrt{(X_1+\frac{1}{2}X_2)^2\tan^2\frac{\pi}{V}+(\frac{1}{2}X_2)^2}$, $b_1=X_1$, $\theta_0=\theta_2=\pi-\arctan[(1+\frac{2X_1}{X_2})\tan\frac{\pi}{V}]$, $\theta_1=2\arctan[(1+\frac{2X_1}{X_2})\tan\frac{\pi}{V}]$, $\psi_0=\arctan[(1+\frac{2X_1}{X_2})\tan\frac{\pi}{V}]-\frac{\pi}{V}$, $\psi_1=\frac{\pi}{V}$, and $\psi_2=\frac{\pi}{2}-\arctan[(1+\frac{2X_1}{X_2})\tan\frac{\pi}{V}]$.

When $l=2,3,\cdots,\frac{M}{V}-1$, $P_{e,l}^{(\mathcal{A})}$ is given by
\begin{align}\label{Pe_star_4}
    P_{e,l}^{(\mathcal{A})}=\frac{1}{2\pi}\sum_{k=0}^{3}\int_0^{\theta_k}\exp
    \Big[-\frac{\rho'b_k^2\sin^2\psi_k}{\sin^2(\theta+\psi_k)}\Big]\text{d}\theta,
\end{align}
where $b_0\!=\!b_3\!=\!\sqrt{\!\big(\!X_1\!+\!\cdots\!+\!X_{l}\!+\!\frac{1}{2}X_{l+1}\!\big)^2
\tan^2\frac{\pi}{V}\!+\!\big(\!\frac{1}{2}X_{l+1}\!\big)^2}$, $b_1=b_2=\sqrt{(X_1+\cdots+X_{l-1}+\frac{1}{2}X_{l})^2\tan^2\frac{\pi}{V}+(\frac{1}{2}X_{l})^2}$, $\theta_0=\theta_2=\pi-\arctan[(1+\frac{2(X_1+\cdots+X_{l-1})}{X_{l}})\tan\frac{\pi}{V}]$, $\theta_1\!\!=\!\!2\arctan[(1+\frac{2(X_1+\cdots+X_{l-1})}{X_{l}})\tan\frac{\pi}{V}]$, $\theta_3\!\!=\!\!2\arctan[(1+\frac{2(X_1+\cdots+X_{l})}{X_{l+1}})\tan\frac{\pi}{V}]$, $\psi_0\!\!=\!\!\arctan[(1+\frac{2(X_1+\cdots+X_{l})}{X_{l+1}})\tan\frac{\pi}{V}]-\frac{\pi}{V}$, $\psi_1\!\!=\!\!\frac{\pi}{2}-\arctan[(1+\frac{2(X_1+\cdots+X_{l-1})}{X_{l}})\tan\frac{\pi}{V}]$, $\psi_2\!\!=\!\!\arctan[(1+\frac{2(X_1+\cdots+X_{l-1})}{X_{l}})\tan\frac{\pi}{V}]+\frac{\pi}{V}$ and $\psi_3\!\!=\!\!\frac{\pi}{2}-\arctan[(1+\frac{2(X_1+\cdots+X_{l})}{X_{l+1}})\tan\frac{\pi}{V}]$.

Furthermore, $P_{e,\frac{M}{V}}^{(\mathcal{A})}$ is given by
\begin{align}\label{Pe_star_6}
    \notag P_{e,\frac{M}{V}}^{(\mathcal{A})}&=\frac{1}{2\pi}\int_0^{\theta_0}\exp
    \Big[-\frac{\rho'b_0^2\sin^2\psi_0}{\sin^2(\theta+\psi_0)}\Big]\text{d}\theta+\\
    &\quad\frac{1}{\pi}\sum_{k=0}^{1}\int_0^{\pi-\psi_1}\exp
    \Big[-\frac{\rho'b_0^2\sin^2\psi_1}{\sin^2\theta}\Big]\text{d}\theta,
\end{align}
where $b_0\!\!=\!\!\sqrt{(X_1\!+\!\cdots\!+\!X_{\frac{M}{V}-1}\!+\!\frac{1}{2}
X_{\frac{M}{V}})^2\tan^2\frac{\pi}{V}\!+\!(\frac{1}{2}X_{\frac{M}{V}})^2}$, $\theta_0\!\!=\!\!2\arctan[(1+\frac{2(X_1+\cdots
+X_{\frac{M}{V}-1})}{X_{\frac{M}{V}}})\tan\frac{\pi}{V}]$, $\psi_0\!\!=\!\!\frac{\pi}{2}-\arctan[(1+\frac{2(X_1
+\cdots+X_{\frac{M}{V}-1})}{X_{\frac{M}{V}}})\tan\frac{\pi}{V}]$ and $\psi_1\!\!=\!\!\arctan[(1+\frac{2(X_1+\cdots+X_{\frac{M}{V}-1})}
{X_{\frac{M}{V}}})\tan\frac{\pi}{V}]+\frac{\pi}{V}$.

Then, upon substituting (\ref{Pe_star_2}), (\ref{Pe_star_4}) and (\ref{Pe_star_6}) into (\ref{Pe_star_1}), we arrive at the theoretical SEP of $\mathcal{A}_{\frac{M}{V}}^{V}$ A-PSK. Since the final result includes the values of $X_1,X_2,\cdots,X_{\frac{M}{V}}$, it can be evaluated numerically.

\subsubsection{Symbol Error Probability of QA-PSK}
In the ML detection, the SEP is determined by the Euclidean distance of the received signal points from the respective decision boundaries, in which the lowest distances play a dominant role. Thus, the SEP of QA-PSK can be calculated by neglecting the effect of high-distance decision boundary. Therefore, observe from Fig. \ref{Fig_2} (c) that the SEP of $\mathcal{Q}_{\frac{M}{V}}^{V}$ QA-PSK, denoted as $P_e^{\mathcal{(Q)}}$, is given by
\begin{align}\label{Pe_diamond}
    \notag P_e^{(\mathcal{Q})}=&\frac{2}{\sqrt{\frac{M}{V}}}\sum_{l=2}^{\sqrt{\frac{M}{V}}}
    \Big[Q\Big(\frac{\sqrt{\rho'}X_l^{\text{(I)}}}{2}\Big)+Q\Big(\frac{\sqrt{\rho'}
    X_l^{\text{(Q)}}}{2}\Big)\Big]+\frac{q}{\frac{M}{V}}\cdot\\
    &\ \sum_{l=2}^{\sqrt{\frac{M}{V}}}
    \Bigg[Q\Big(\sqrt{\frac{\rho'((X_l^{\text{(I)}})^2+(X_l^{\text{(Q)}})^2-2\cos\frac{2\pi}{V}
    X_l^{\text{(I)}}X_l^{\text{(Q)}})}{2}}\Big)\Bigg].
\end{align}
where $Q(\cdot)$ represents the Gaussian Q-function \cite{simon2005digital}, and the constant $q=4$ when $\log_2V=2$ and $q=2$ when $\log_2V\geq3$. Since the final result includes the values of $X_2^{\text{(I)}},X_3^{\text{(I)}},\cdots,X_{\sqrt{\frac{M}{V}}}^{\text{(I)}}$ and $X_2^{\text{(Q)}},X_3^{\text{(Q)}},\cdots,X_{\sqrt{\frac{M}{V}}}^{\text{(Q)}}$, it can be evaluated numerically.

\section{Simulation Results}
In this section, we analyze the performance of the proposed schemes in terms of their DCMC capacity and SEP, against the SoA RIS-based PSK modulation, where the distance between adjacent RIS elements is $\frac{\lambda}{2}$, and the Rician factor is $\kappa=0\text{dB}$. For fairness of comparison with the SoA RIS-based PSK scheme, we assume the number of RAs at the user is $K=1$.

Fig. \ref{Fig_sepctral_efficiency} compares the DCMC capacity $R$ versus the received SNR $\rho$ for the SoA RIS-based 128-PSK, as well as the proposed $\mathcal{A}_{16}^8$ A-PSK and the $\mathcal{Q}_{16}^8$ QA-PSK for different number of RIS elements $N$, where the RIS phase shift resolution is $B=3$ bits. The theoretical (theo.) UB is very tight compared to the simulation results (simu.) for our proposed schemes. It is shown in Fig. \ref{Fig_sepctral_efficiency} that the DCMC capacity reaches a maximum of 7 bit/s/Hz since the modulation order is $M=128$. In the low-SNR region, the DCMC capacity of the SoA RIS-based PSK modulation is higher than that of our proposed A-PSK and QA-PSK scheme. However, the DCMC capacity of our proposed A-PSK and QA-PSK is better than that of the SoA RIS-based PSK scheme, when the receive SNR is higher than $-35\mathrm{dB}$, $-40\mathrm{dB}$ and $-45\mathrm{dB}$ for $N=512$, $N=1024$ and $N=2048$, respectively.

Fig. \ref{Fig_simulation} (a) compares the SEP $P_e$ versus receive SNR $\rho$ for the SoA RIS-based 128-PSK modulation, the proposed $\mathcal{A}_{16}^8$ A-PSK and $\mathcal{Q}_{16}^8$ QA-PSK, with the parameters being the same as in Fig. \ref{Fig_sepctral_efficiency}. This shows that doubling the number of RIS elements yields approximately $6\text{dB}$ gain, since it is proportional to the square of the number of RIS elements $N$. Furthermore, this shows that under the same transmission rate of 7 bit/s/Hz, the $\mathcal{A}_{16}^8$ A-PSK and $\mathcal{Q}_{16}^8$ QA-PSK have better SEP than the RIS-based 128-PSK. Furthermore, QA-PSK outperforms A-PSK, since the transmit signals of QA-PSK are distributed more uniformly than those of A-PSK, which results in higher minimum Euclidean distance in the received signal constellation. It also shows that the theoretical analysis and the simulation results of the QA-PSK scheme match tightly in the high-SNR region. This is due to the fact that in our theoretical analysis, the SEP is derived based on the lowest distances from the received signal points to the respective decision boundaries, which results in tight approximation in the high-SNR region.

Fig. \ref{Fig_simulation} (b) compares the SEP $P_e$ versus the receive SNR $\rho$ of the SoA RIS-based 128-PSK modulation, of the proposed A-PSK and QA-PSK at different transmission rates $R$, for $N=1024$ RIS elements, and for $B=3$ bits. In the SoA RIS-based scheme, 32-PSK, 128-PSK and 512-PSK are employed at the transmission rates of $R=5,7,9$ bit/s/Hz, respectively. By contrast, in our proposed methods, the $\mathcal{A}_{4}^8$, $\mathcal{A}_{16}^8$, $\mathcal{A}_{64}^8$ A-PSK schemes and $\mathcal{Q}_{4}^8$, $\mathcal{Q}_{16}^8$, $\mathcal{Q}_{64}^8$ QA-PSK schemes are employed correspondingly. Observe that at low rates of say $R=5$ bit/s/Hz, the advantage of QA-PSK is not obvious, but at high rates of say $R=9$ bit/s/Hz, QA-PSK considerably outperforms both the SoA RIS-based PSK and A-PSK. Explicitly, our proposed QA-PSK scheme has improved the SEP, especially at high transmission rates.

Fig. \ref{Fig_simulation} (c) compares the SEP $P_e$ versus received SNR $\rho$ of the SoA RIS-based 128-PSK, of the proposed $\mathcal{A}_{16}^8$ A-PSK and $\mathcal{Q}_{16}^8$ QA-PSK at different values of $B$, where the number of RIS elements is $N=64$. As expected, the finite phase shift resolution degrades the SEP of the SoA RIS-based PSK, but it has little effect on our proposed schemes.

\section{Conclusions}
The novel A-PSK and QA-PSK schemes are proposed for RIS-based transmitters, where the signals of the multi-RA receiver were coherently combined based on the statistical CSI, when using the ML detection method. Both our theoretical analysis and simulation results show that the proposed schemes outperform the SoA RIS-based PSK modulation in terms of both the DCMC capacity and SEP, especially for high rate transmission and finite RIS phase shift resolution.
\bibliographystyle{IEEEtran}
\bibliography{IEEEabrv,TAMS}
\end{document}